\documentclass[12pt]{article}

\begin{document}
\pagestyle{empty} 
\def\d{{\delta }}
\par

\par

Aug. 21, 2003.

To: JIIP editor, Professor M.M.Lavrentiev, 
Applicable Analysis editor, Professor R. P. Gilbert

A brief information to the reader: I have received from
the editor of the Journal of Inverse and Ill-Posed
Problems (JIIP), Prof. M.M.Lavrentiev. in March, 2003, a 
letter of P.Sabatier to the editor of JIIP.
In this letter Sabatier wrote that
my statement {\bf "The Newton-Sabatier (NS) method for inverting the 
fixed-energy phase shifts for a potential is fundamentally wrong"}
is erroneous. This statement appeared in [3]. I have proposed to the 
editor of JIIP to publish Sabatier's letter and my reply.
This reply the reader can find below.
The editor of JIIP informed me that Sabatier's letter and my reply will
not be published in JIIP.

However, I have found out on Aug.17, 2003, that Sabatier's letter was 
published in Applicable Analysis, 82, N3, (2003), 399-400.
The editor of Applicable Analysis, Prof. R. P. Gilbert,
did not send me Sabatier's letter and I was not given a chance to reply to
this letter. Below the reader can find my comments on the published letter 
of P.Sabatier.

Comments on the letter of P.Sabatier, An erroneous statement, 
Applicable Analysis, 82, N3, (2003), 399-400.

The published letter of Professor Sabatier is an edited version of his
letter,
sent to me in March, 2003, by the JIIP editor, and my reply is appended
below.

The editor of Applicable Analysis did not send me Sabatier's
letter, so I was not able to reply to this letter right away.

In his letter Sabatier writes that my statement 
{\bf "The Newton-Sabatier (NS) method for inverting the fixed-energy
phase shifts for a potential is fundamentally wrong"}
is erroneous.

In [2] my statement is proved. It is explained in [2] that the
foundations of the NS method are wrong, and, in particular, that the basic
integral equation of this method, in general is not uniquely solvable for
some $r>0$, contrary to the implicit assumption of the NS method.
If this equation is not uniquely solvable for some $r>0$, then the 
NS method breaks down: it produces a potential which is not
locally integrable.
 
Thus, my conclusion was: {\it the NS theory as an inverse scattering
theory is fundamentally wrong in the sense that its foundations are
wrong}. This conclusion is
fully justified in [2].

Sabatier does not discuss this main issue. 
In fact he does not discuss any of my concrete statements, given in [2].

In [2] I have discussed the following questions:

a) Is the NS theory a valid inverse scattering
theory? 

and

b) Are the foundations of the NS theory correct?

I gave the answer NO to both questions, and justified my answers.

Sabatier does not argue with my justification of these answers.
The validity
of the NS theory as an inverse scattering theory is not even mentioned
in his letter.
 
Instead, he tries to create an impression that the NS method can serve,
under some conditions, as a parameter-fitting method, which
produces some potential, given the set of
"non-exceptional" phase shifts. The question of whether this 
potential generates the original phase
shifts is not dicussed in Sabatier's letter. In [1, p.205],
there is a discussion of this basic question, but the proof of the 
"consistency" of the NS method is not correct.

 I have demonstrated in [2] that the NS method cannot serve as a
parameter-fitting method, because, in general, the basic integral
equation is not uniquely solvable for some $r>0$, and in this case the NS
method breaks down. I do not want to go into further discussion of the 
errors in the presentation of the NS method in [1]. However, one wrong
statement,
which is given as a conclusion of the "proof of the consistency" of the NS
method in [1, p.205], may be of interest to the reader, and its
discussion is short.
Sabatier claims in [1, p.205], that the NS method gives "one (only one)
potential which decreases faster than $r^{-\frac 32}$."
This statement is incorrect: take any compactly supported, integrable, 
real-valued potential $q$, construct its phase shifts at a fixed positive
energy, and apply the NS method for the reconstruction
If the NS method 
produces a potential $q_{NS}$, decaying faster than $r^{-\frac 32}$, and
if this potential generates the 
original phase shifts, then one has two potentials, $q$ and $q_{NS}$,
which both decay faster than  $r^{-\frac 32}$, and which are different,
because the NS method cannot produce a compactly supported potential
([2]). Thus, Sabatier's statement in [1, p.205] is incorrect.
Since Sabatier claims in his letter that the NS method should work for 
"almost all" phase shifts (the meaning of "almost all" is not specified),
one may hope that the phase shifts, generated by some compactly
supported potential, are not "exceptional", because the set
of compactly supported integrable potentials  is dense
in many sets of integrable
potentials.

The other statements in Sabatier's letter are discussed in detail in the
appended reply to his original letter of March, 2003.

{\it To summarize:} my statement that {\bf "the NS inversion theory is
fundamentally
wrong, in the sense that its foundations are wrong"}  is {\it correct}. 

Alexander Ramm 
\vfill

====================
March 14, 2003.

To: JIIP editor, Professor M.M.Lavrent'ev,  
Applicable Analysis editor, Professor R. P. Gilbert

Some of the conclusions proved in [2] are:

1. In the NS theory R.Newton and P.Sabatier assume {\it without proof} 
that equation (12.1.2) 
in 
[1], with $f$ defined in (12.2.1) in [1], is uniquely solvable for all 
$r>0$.
This assumption is fundamentally wrong. In general, this equation
is not uniquely solvable for some $r>0$ and, in this case, the 
NS method breaks down: it produces a potential which is not locally 
integrable.

2. In the NS theory R.Newton and P.Sabatier  assume {\it without proof} 
that the 
transformation 
kernel can be represented by formula (12.2.3) from [1]. This assumption is  
fundamentally wrong, because, in general, the transformation kernel
does not have the form (12.2.3) from [1].
In [1],  the existence and uniqueness of the transformation kernel,
used there, is not proved, and its properties are not studied.
A proof of the existence and uniqueness of it is given in [6],
where the properties of the transformation kernel, used in [1],
are studied.

 The above assumptions form the basis of the NS inversion theory.
{\it Therefore the NS theory is fundamentally wrong; that is, its basic 
assumptions are wrong.}  

This theory is {\it not} 
an inversion method for solving the inverse scattering problem.
An inversion method is a method that allows one to recover
a potential $q$, given 
the scattering data (fixed-energy phase shifts in our case) generated
by this  $q$. For the NS procedure, (even if it can 
be carried through), it is not proved that the potential it produces 
generates the original scattering data (the proof given in [1], p.205, is 
not convincing [2]).

   The above argument does not depend on the class of potentials 
in which the solution to the inverse scattering problem is sought.
Therefore, Professor Sabatier's remark about "favorite 
classes of potentials"
is irrelevant and shows that he has missed the main point.
 
Professor Sabatier writes that the inverse scattering 
problem is to find $c_\ell,$ given the phase shifts $\d_\ell$. This is
incorrect: the problem is to find a potential $q$ from the corresponding 
phase shifts. 
In [2] the questions related to finding $c_\ell$ from $\d_\ell$
are not discussed.

 Professor Sabatier writes that he "does not object against Ramm 
republishing 40 years old results". Apparently Professor Sabatier
has missed the basic point: I have proved that {\it 
the 40 year-old NS theory
is fundamentally wrong}, and presented a detailed justification of this 
conclusion.

 Professor Sabatier mentions that the uniqueness theorem for the inverse 
scattering 
problem in the class of compactly supported potentials belongs to Loeffel, 
and, he continues: " Ramm claims that he proved the theorem (1987)".
 This statement of Professor Sabatier is erroneous:
whereas Loeffel dealt with 1D case,
namely, with the spherically-symmetric potentials,
 my result,
mentioned in [2], is the uniqueness theorem for the 3D inverse scattering 
problem with fixed-energy data.
My uniqueness theorem implies the uniqueness theorem for 
the case of the spherically-symmetric compactly supported potentials.

Professor Sabatier writes "there are many potentials fitting the same set 
of phase shifts". It is not clear what he means: if "fitting"
means that two compactly supported potentials, $q_1$ and $q_2$, produce 
identical fixed-energy phase shifts for all $\ell$, then Professor Sabatier's
statement is incorrect, and $q_1=q_2$. If "fitting" means
approximately equal (in some sense) then
  Professor Sabatier's statement is true if it is taken out of the 
context; however, in 
the context it shows that he does not understand the difference between 
the
notion of inverse scattering theory and the notion of ill-posedness
of the inverse scattering problem with fixed-energy data. An inverse 
scattering theory is the theory that reconstructs a potential $q,$
which belongs to some class of potentials, from the scattering data, 
generated by this potential.
This problem may be ill-posed in the sense that $q$ does not depend 
continuously on the data. A problem may have a unique solution and be 
ill-posed. The inverse scattering theory may give results of the 
following nature: given the scattering data, generated by a potential $q$
from some class, one can uniquely reconstruct this $q$. 
Also, it may give an algorithm for finding $q$ from the scattering 
data.
However, there can be many different potentials (even in the same class,
but, also in other classes as well),
which produce the scattering data close in some sense to the 
original data. If $q$ does not depend continuously
(in some sense) on the data, then the inverse scattering 
problem is ill-posed.
Theoretical investigation of the uniqueness of the solution to the inverse 
scattering problem and finding a reconstruction formula
or algorithm for finding $q$ from the exact data are questions that 
are {\it separate} from the numerical reconstruction given noisy data and 
from the related notion of ill-posedness. All this is well known;
I am stating this, because Professor Sabatier confuses the issues
in his letter.  

 Professor Sabatier writes that
the NS inversion procedure works in "most cases".
Exactly the opposite is true: {\it the NS procedure does not work in most 
cases}, because in most cases equation (12.1.2) in [1] does not have
a unique solution for some $r>0$. This was explained in [2].
In [4] there was an attempt to prove that a modified version of equation 
(12.1.2) in [1] has a unique solution for all $r>0$, but the claim
in [4] is erroneous: a counterexample  to this claim is 
constructed in [5].

Professor Sabatier writes that Fredholm's integral equation with 
spectral parameter is "not generically solvable in this sense".
This is incorrect: such an equation has a discrete set of
eigenvalues and therefore, by the Fredholm alternative, it is uniquely 
solvable "generically"- that is, for all values of the spectral parameter, 
except for a countable set (of Lebesgue's measure zero). 
 
 Professor  Sabatier writes that a student can work out
"one or two parameters examples...". This is true: these examples 
of solving linear Fredholm integral equations with degenerate 
kernels are trivial. However, these examples have nothing to 
do with the inverse scattering problem. {\it They are not examples of 
finding a potential from the 
given phase shifts.}


 It follows from the above, that the claim in [3] stating that the NS 
inversion theory is fundamentally wrong, is correct and justified 
completely and in detail in [2].

 I do not find the last paragraph of Professor Sabatier's letter  
worthy of discussion.

  One may argue that the NS theory, being fundamentally wrong as an 
inverse scattering theory, can be used as a parameter-fitting procedure.
However, it is proved in [2] that the set of potentials, which can 
possibly be obtained by the NS procedure, is not dense in
the standard class $L_{1,1}$ of potentials $q$ in the norm $||q||:=
\int_0^\infty x|q(x)|dx$, for example. Therefore, if one wishes to solve 
the 
inverse scattering problem with the given fixed-energy phase shifts,
corresponding to a potential in  $L_{1,1}$,
one cannot do this using the NS procedure as a parameter-fitting
procedure, in general.

In the literature there are papers with examples 
showing that although  the original NS procedure does not work 
numerically, its modification, proposed by Professor Scheid and his 
colleagues, can be used as a parameter-fitting procedure in some
cases. This, however, is the point which was not discussed in [2],
where I have analyzed the foundations of the NS theory.

References:
[1] Chadan K., Sabatier P.,
{\sl Inverse Problems in Quantum Scattering Theory}, Springer, New York, 
1989.

[2] Ramm, A.G.,Analysis of the Newton-Sabatier scheme for inverting
fixed-energy phase shifts, Applic. Analysis, 81, N4, (2002), 965-975.

[3] Gutman, S., Ramm A.G., Scheid W., Inverse scattering by the stability 
index method,
Jour. of Inverse and Ill-Posed Probl., 10, N5, (2002), 487-502.

[4] J.Cox and K.Thompson, Note on the uniqueness of the
solution of an equation of interest
in inverse scattering problem,  J. Math.Phys.,  11, N3,
(1970), 815-817.

[5]  Ramm A.G.,  A counterexample to a uniqueness result,
Applic. Analysis, 81, N4, (2002), 833-836.

[6]  Ramm A.G., Inverse scattering problem with part of the 
fixed-energy
phase shifts, Comm. Math. Phys.  207, N1, (1999), 231-247.

\vspace{.25in}

Alexander Ramm \\
ramm@math.ksu.edu
\end{document}